\documentclass[prb,aps,twocolumn,floatfix,showpacs]{revtex4}
\usepackage{graphicx,graphics,psfrag,amsmath,calc}
\usepackage{epsfig}
\begin{document}
\title{Electrostatically induced phase transitions in superconducting complex oxides}
\author{Li Han and C. A. R. S{\'a} de Melo}
\affiliation{School of Physics, Georgia Institute of Technology, Atlanta, Georgia 30332, USA}
\date{\today}

\begin{abstract}

We describe quantum phase transitions in superconducting complex oxides which
could be tuned by electrostatic charge transfer. Using a simple model for
the superconductivity of a thin film or surface of a bulk copper oxide, we show that
tuning the carrier density may allow the visitation of several superconducting phases with
different pairing symmetries such as extended $s$- $(se)$, $d$- and
$(se \pm id)$-wave. We construct a universal phase diagram for
single-band superconductors with $se$- and $d$-wave components of the order parameter
based on symmetry considerations alone. For a specific model with nearest neighbor
attraction, we obtain the phase diagram in the interaction versus filling factor space
showing the boundaries of the possible phases. Finally, we calculate the superfluid density
and penetration depth as characteristic properties of each phase.

\end{abstract}
\maketitle

%
%

Substantial progress in tuning carrier density has been achieved
recently in systems like amorphous Bismuth, where a
superconducting-insulator transition was driven by electrostatic
fields~\cite{goldman-2005}. This experimental realization has opened
the possibility of studying quantum phase transitions by tuning
carrier density electrostatically. Several groups are currently
attempting the use of electrostatic doping to control the carrier
density in complex oxides, with particular attention to cuprate
superconductors. The phase diagram of cuprate superconductors as a
function of chemical substitution (oxygen deficiency) is known for
Lanthanum (Yttrium) cuprates, but as it is well documented in the
literature that the relation between chemical doping or oxygen
deficiency to carrier density is very complicated~\cite{ahn-2006}.

Eletrostatic doping has substantial advantages over chemical doping,
as the carrier density can be tuned continuously and does not
produce undesired effects such as disorder or changes in the crystal
structure as chemical doping or oxygen deficiency often does. The
method when applied to cuprates should be able to answer the long
standing question of what really is the phase diagram of the system
as a function of carrier density, and to clarify the origin of the
dome that reflects an optimally doped superconductor with the
highest critical temperature. In particular, if the results of
electrostatic doping are substantially different from those of
chemical doping, showing for instance that the critical temperature
continues to grow beyond the expected optimal chemical doping point,
then it could be directly inferred that chemical doping does
introduce undesired effects.

Although electrostatic doping has remarkable advantages over chemical doping, its use
is limited to thin films and the surface of bulk systems. However, it is probably
the best way to study the dependence of physical properties of complex oxides and other
materials on carrier density. Furthermore, from the point of view of devices,
the electrostatic tuning of the phase diagram at fixed temperature would be extremely
useful to produce a superconducting switch, as the resistance of a thin film could be turned
on and off via a gate voltage.

It is in anticipation of these exciting experiments currently underway in
several labs around the world,
that we describe the possibility of using
electrostatic tuning of carrier density to study quantum phase transitions
in complex oxides. As an example we chose to study a simple model of complex
oxides representing a cuprate superconductor. We model such a system via
an extended attractive Hubbard model in a two-dimensional lattice, and
derive the universal phase diagram when only two pure order parameter symmetries
are allowed. Based on symmetry alone there are several possibilities:
pure extended $s$- $(se)$ and $d$-wave phases, and
$(se \pm d)$ phases that do not break time-reversal symmetry and $(se \pm id)$ phases that do.
For a specific model where only nearest neighbor attraction is
included, the $se$-wave phase dominates at lower filling factor, while the $d$-wave solution
dominates at higher filling factor, with a $(se \pm id)$-wave phase in between, whereas
the $(se \pm d)$-wave is not accessible.
The phase diagram for such a minimal model is very rich, and the existence of various
phases as a function of filling factor for fixed interactions can be tested via measurements of
the penetration depth (superfluid density) at low temperatures.

To study the physics discussed above we start from a two-dimensional hamiltonian
\begin{equation}
\label{eqn:hamiltonian}
H = -t\sum_{\langle ij \rangle \sigma} c^{\dagger}_{i \sigma} c_{j \sigma}
- U \sum_i n_{i \uparrow} n_{i \downarrow}
- V \sum_{\langle ij \rangle \sigma \sigma^\prime}
n_{i \sigma} n_{j \sigma^\prime}
\end{equation}
describing thin films or the surface layer of a bulk cuprate
on a square lattice. Here, the local particle number operator
is $n_{i \sigma} = c^\dagger_{i \sigma} c_{i \sigma}$, while $-U$ and $-V$
correspond to on-site and nearest neighbor attractions, respectively.

In order to establish the quantum phases as a function of
carrier density, we start by constructing the partition function
$Z = \int {\cal D} c^\dagger {\cal D} c e^{S}$ for the action
\begin{equation}
\label{eqn:action} S = \int_0^{\beta} d\tau \left[ \sum_{i\sigma}
c^{\dagger}_{i\sigma} (\tau) (-\partial_\tau + \mu) c_{i\sigma}
(\tau) - H (c^\dagger, c) \right]
\end{equation}

Introducing order parameters for local $s$-wave $\tilde
\Delta_{sl}$, extended $s$-wave $\tilde \Delta_{se}$ and $d$-wave
$\tilde \Delta_{d}$ pairing, we can write the action as
\begin{equation}
\label{eqn:action-order-parameter} S = - \frac{N_s}{T} \sum_{q,
\alpha} \frac{\vert \tilde \Delta_{\alpha} (q) \vert^2}{V_{\alpha}}
+ \mathrm{Tr} \ln \left( \frac {{\bf G}_{0}^{-1}}{T} - \frac{ {\bf
V}}{T} \right) + \frac{\mu N_s}{T},
\end{equation}
where $\alpha = sl, se, d$; the interactions are
$V_{sl} = U$, $V_{se} = V_{d} = V$; and the four-vector $q = (i\nu_n, {\bf q})$.

The inverse free fermion propagator matrix is
\begin{equation}
{\bf G}_{0}^{-1} (k, k^\prime) =
\left(
\begin{array}{cc}
 i \omega_n - \xi_{\bf k}  &   0 \\
0 & i \omega_n + \xi_{\bf k}
\end{array}
\right) \delta_{k, k^\prime}
\end{equation}
with kinetic energy $\xi_{\bf k} = \epsilon_{\bf k} - \mu$,
band dispersion $\epsilon_{\bf k} = - 2t \left[ \cos(k_x a) + \cos(k_y a) \right]$,
chemical potential $\mu$, four-vector $ k = (i\omega_n, {\bf k})$,
and unit cell length $a$.
The additional matrix appearing in Eq.~(\ref{eqn:action-order-parameter}) is
\begin{equation}
\label{eqn:interaction-matrix}
{\bf V} (k, k^\prime) =
\left(
\begin{array}{cc}
0 &  {\widetilde \Delta}_{\alpha} (k - k^\prime) \\
{\widetilde \Delta}_{\alpha}^{*} (-k + k^\prime) & 0
\end{array}
\right) \lambda_{\alpha}({\bf k}, {\bf k}^\prime ) ,
\end{equation}
where the Einstein summation over $\alpha$ is understood,
and $\lambda_{\alpha}({\bf k}, {\bf k}^\prime )$ are the
symmetry factors for the order parameters, which in the
limit of zero momentum pairing $({\bf k} = {\bf k}^\prime )$ become
$\lambda_{sl} ({\bf k}, {\bf k}) = 1$,
$\lambda_{se}({\bf k}, {\bf k}) = \cos(k_x a) + \cos(k_y a)$,
$\lambda_{d}({\bf k}, {\bf k}) = \cos(k_x a) - \cos(k_y a)$.

The saddle point approximation gives qualitatively correct
results either at low temperatures $(T \ll T_c)$ for any interaction
strength or for weak interactions $V \ll bw = 8t$ at
any temperature~\cite{nsr-1985, sademelo-2005}, where $bw$ is the bandwidth
in two-dimensions.
In these cases, the effective action can be
approximated by
\begin{equation}
\label{eqn:action-saddle-point} S = - \frac{N_s}{UT} \vert {\tilde
\Delta}_{sl} \vert^2 - \frac{N_s}{VT} ( \vert {\tilde \Delta}_{se}
\vert^2 + \vert {\tilde \Delta}_{d} \vert^2  ) + S_2 + \frac{\mu
N_s}{T}.
\end{equation}
The second term in the action
$S_2 = \sum_{{\bf k},\gamma} \ln \left[ 1 + \exp(-E_{ {\bf k},\gamma }/T ) \right]$,
contains the quasiparticle ($\gamma = 2$) or quasihole ($\gamma = 1$)
energies
$E_{{\bf k}, \gamma} =
(-)^\gamma
\sqrt{
\xi_{\bf k}^2 +
\vert {\tilde \Delta}_\alpha \lambda_{\alpha} ({\bf k}) \vert^2
},
$
where the symmetry function is defined as
$\lambda_{\alpha} ({\bf k}) = \lambda_{\alpha} ({\bf k}, {\bf k})$.

Notice that even in such a simple model with local $(-U)$ and
nearest neighbor $(-V)$ attractions the number of possible phases is
quite large. There are three possible pure phases: local $s$-wave
($sl$); extended $s$-wave ($se$) and $d$-wave ($d$). There are
several possible binary mixed phases $sl \pm se$, $sl \pm d$, and
$se \pm d$, which do not break time-reversal symmetry, as well as
those that do like $sl \pm i se$, $sl \pm i d$, and $se \pm id$.
Lastly there are also several ternary mixed phases involving all
three symmetries $sl$, $se$, and $d$. Since the situation is quite
complicated in the more general case, for clarity and simplicity, we
study here the case of $U = 0$, which allows for solutions involving
only $se$- and $d$-wave symmetries. In this simpler case there are
only six possible phases. The pure phases are $se$ and $d$. The
mixed phases are $se \pm d$, and $se \pm i d$. Thus, from now on, we
will confine ourselves to this simpler analysis. In this case, the
order parameter equations can be obtained by minimizing the action
$S$ through the conditions $\delta S/ \delta {\tilde \Delta}_{se}^*
= 0$ and $\delta S/ \delta {\tilde \Delta}_{d}^*  = 0$. For the
$se$-wave component of the order parameter the first condition leads
to
\begin{equation}
\label{eqn:order-parameter-s-wave}
\tilde \Delta_{se} =
\frac{V}{N_s}
\sum_{\bf k}
\frac{\tanh (E_{{\bf k}, 2}/2T)} {2  E_{{\bf k}, 2} }
\Lambda_{se} ({\bf k}),
\end{equation}
where
$
\Lambda_{se} ({\bf k}) =
\tilde \Delta_{se} \left[ \lambda_{se} ({\bf k}) \right]^2
+
\tilde \Delta_d  \lambda_{se} ({\bf k}) \lambda_d ({\bf k}).
$
Correspondingly for the $d$-wave component
\begin{equation}
\label{eqn:order-parameter-d-wave}
\tilde \Delta_{d} =
\frac{V}{N_s}
\sum_{\bf k}
\frac{\tanh (E_{{\bf k}, 2}/2T)} {2  E_{{\bf k}, 2} }
\Lambda_{d} ({\bf k}),
\end{equation}
where
$
\Lambda_{d} ({\bf k}) =
\tilde \Delta_{d} \left[ \lambda_d ({\bf k}) \right]^2
+
\tilde \Delta_{se}  \lambda_{se} ({\bf k}) \lambda_d ({\bf k}).
$

The number equation that fixes the chemical potential is obtained
through the thermodynamic relation $N = -\partial \Omega / \partial \mu$,
where $\Omega = - T \ln Z$ is the thermodynamic potential. In the
present approximation the thermodynamic potential is
$\Omega = -T S$, and the number equation reduces to
\begin{equation}
\label{eqn:number}
\nu  = \frac{1}{N_s} \sum_{\bf k}
\left[
1 - \frac{\xi_{\bf k}}{E_{{\bf k}, 2}}
\tanh ( E_{{\bf k}, 2}/2T )
\right],
\end{equation}
where $\nu = N/N_s$ is the filling factor.

The phase difference between the $s$-wave and $d$-wave components of
the order parameter is defined to be $\delta \phi = \phi_d -
\phi_{se}$, where $\phi_d$ is the phase of the $d$-wave order
parameter ${\tilde \Delta}_d = \vert {\tilde \Delta}_d \vert
e^{i\phi_d}$ and $\phi_{se}$ is the phase of the $s$-wave order
parameter ${\tilde \Delta}_{se} = \vert {\tilde \Delta}_{se} \vert
e^{i\phi_{se}}$. Simultaneous solutions of
Eqs.~(\ref{eqn:order-parameter-s-wave}),~(\ref{eqn:order-parameter-d-wave}),
and~(\ref{eqn:number}) reduce to saddle-point solutions of $\Omega$
only for $\delta \phi = \pi/2, 3\pi/2$ which correspond to $se \pm i
d$ phases that break time-reversal symmetry, and $\delta \phi = 0,
\pi$ which correspond to $se \pm d$ phases that do not.

%
%

\begin{figure} [htb]
\psfrag {?} {$\pm$}
\centerline{\scalebox{0.50}{\includegraphics{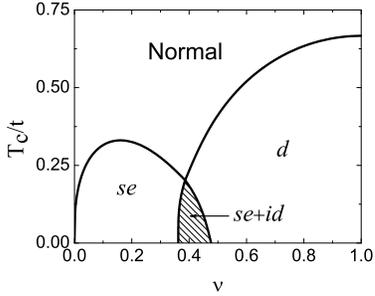} } }
\caption{ \label{fig:critical-temperature} Plots of the critical
temperature $T_c$ as a function of filling factor $\nu$ at fixed
interaction $(V/t = 3.0)$, showing the the $se$-, $d$-, and $(se +
id)$-wave phases. Notice the tetracritical point where the normal
and all superconducting phases meet. }
\end{figure}

The saddle point critical temperature can be obtained by setting the
order parameters $\tilde \Delta_{se} = 0$ and $\tilde \Delta_{d} = 0
$ in
Eqs.~(\ref{eqn:order-parameter-s-wave}),~(\ref{eqn:order-parameter-d-wave}),
and~(\ref{eqn:number}). In this case, the filling factor dependence
of critical temperature $T_c (\nu)$ and critical chemical potential
$\mu_c (\nu)$ can be obtained for pure $se$- and $d$-wave
symmetries. The solutions for $T_c (\nu)$ are shown in
Fig.~\ref{fig:critical-temperature} for $V/t = 3.0$. Notice that the
$s$-wave phase is favored at lower filling factors, while the
$d$-wave phase is favored at higher filling factors. In addition,
the critical temperature is symmetric about $\nu = 1$, since the
Helmholtz free energy $F = \Omega + \mu N$ is invariant under the
transformation $\mu \to - \mu$ and $\nu \to 2 - \nu$.

%

%
%

\begin{figure} [htb]
\psfrag {+} {$\pm$}
\centerline{\scalebox{0.50}{\includegraphics{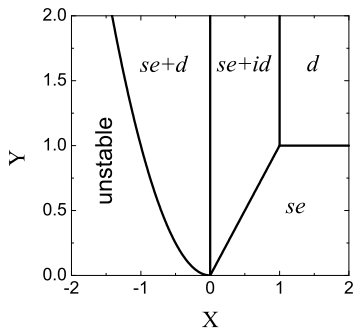} } }
\caption{ \label{fig:universal-phase-diagram} Universal phase
diagram for superconductors exhibiting order parameters with $se$-
and $d$-wave components is shown, in the space of the dimensionless
parameters $X = \lbrace b_{sd} \left[ 1 + \frac12 \cos (2 \delta
\phi ) \right] / b_d  \rbrace \times \vert a_d / a_{se} \vert$ and
$Y = ( b_{se} / b_d ) \times \vert a_d / a_{se} \vert^2$. Depending
on microscopic details, different regions of the phase diagram are
accessible. }
\end{figure}

In order to construct a universal phase diagram for all possible
phases it is important to construct the Ginzburg-Landau theory near
$T_c$ by expanding the action described in
Eq.~(\ref{eqn:action-saddle-point}) in terms of the order parameters
$\tilde \Delta_{se}$, $\tilde \Delta_d$ and their complex
conjugates. From the thermodynamic potential $\Omega = - TS$, we can
calculate the Helmholtz free energy $F = \Omega + \mu N$. The free
energy per site ${\cal F} = F/N_s$ takes the simple form
\begin{equation}
\label{eqn:free-energy-near-critical-temperature}
\begin{array}{c}
{\cal F} =
a_{se} \vert \tilde \Delta_{se} \vert^2 +
a_d \vert \tilde \Delta_d \vert^2 +
b_{se} \vert \tilde \Delta_{se} \vert^4 +
b_d \vert \tilde \Delta_d \vert^4 + \\
2 b_{sd} \left[ 1 + \frac {1}{2} \cos (2 \delta \phi) \right]
\vert \tilde \Delta_{se} \vert^2  \vert \tilde \Delta_d \vert^2
+ \mu (\nu - 1).
\end{array}
\end{equation}
This expression has precisely the expected form, based on symmetry
grounds alone, when paring with $se$- and $d$-wave components are
possible. However, the coefficients $a$ and $b$ depend explicitly on
the parameters of the microscopic model used. For the construction
of the universal phase diagram we introduce the dimensionless
parameters $X = \lbrace b_{sd} \left[ 1 + \frac12 \cos (2 \delta
\phi ) \right] / b_d  \rbrace \times \vert a_d / a_{se} \vert$, and
$Y = ( b_{se} / b_d ) \times \vert a_d / a_{se} \vert^2$.
Minimization of the free energy in
Eq.~(\ref{eqn:free-energy-near-critical-temperature}) and a
stability analysis leads to the universal phase diagram shown in
Fig.~\ref{fig:universal-phase-diagram}, where all possible phases
($se$, $d$, $se \pm d$ and $se \pm id$) are indicated. Notice that
the free energy depends only on $2 \delta \phi$ and does not
distinguish between the phases $se + d$ and $se - d$, which are thus
degenerate. The same applies to the phases $se + id$ and $se - id$,
which are also degenerate. In the particular case of $se \pm id$
phases, time-reversal symmetry is broken while chirality is not, the
latter of which requires additional terms in the free energy for the
distinction between the $se + id$ and $se - id$ phases. Notice that
a tetracritical point occurs at $Y = X = 1$, where the normal and
superconducting phases with $se$, $d$ and $se \pm id$ symmetries
meet.

\begin{figure} [htb]
\psfrag {+} {$\pm$}
\centerline{\scalebox{0.50}{\includegraphics{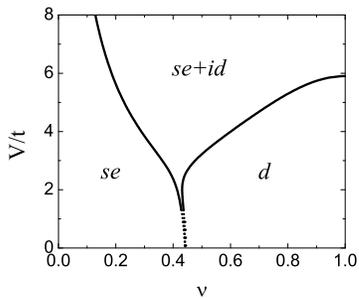} } }
\caption{ \label{fig:zero-temperature-phase-diagram} The
zero-temperature phase diagram in the interaction $V/t$ versus
filling factor $\nu$ plane showing the boundaries between all
accessible phases $se$, $d$, and $se + id$ for the microscopic model
discussed. The phase transitions are always continuous across the
boundaries. }
\end{figure}

For the specific microscopic model discussed above the variable $X$
is always positive, such that the system is always stable and the
$se \pm d$ phases are not accessible. Thus, we elaborate further
only on the accessible phases $se$, $d$, and $se \pm id$. At low
temperatures $T \ll T_c$, the Helmholtz free energy per site can be
approximated by
\begin{equation}
{\cal F} = \frac{1}{V} \sum_{\alpha} \vert \tilde \Delta_{\alpha}
\vert^2 - \frac{T}{N_s} \sum_{{\bf k}, \gamma} \ln \left[ 1 + \exp(-
E_{ {\bf k}, \gamma}/T ) \right] + \mu (\nu - 1) \nonumber
\end{equation}
%
%
%
for any interaction strength $V/t$. A comparison of the free
energies for all the accessible phases produces the phase diagram in
the interaction $(V/t)$ versus filling factor $(\nu)$ space shown in
Fig.~\ref{fig:zero-temperature-phase-diagram}. Notice that pure $se$
and $d$ phases are always separated by $(se \pm id)$ phases, and
their phase boundaries describe continuous transitions. For weak
attractions the region of filling factors where $(se \pm id)$ phase
is realized is very narrow. However, the region increases
substantially as the interaction $V/t$ gets larger. For instance
when $V/t = 3.0$, the $(se \pm id)$ phase exists between $\nu_{min}
\approx 0.36$ and $\nu_{max} \approx 0.48$.

We had hoped that a topological quantum phase transition~\cite{sademelo-2000}
characterized by the emergence of a gapfull $d$-wave superconductor from
a gapless $d$-wave superconductor would also emerge within the $d$-wave region
of the phase diagram. However, within the $d$-wave boundary the chemical potential
always fall inside the band limits $(\vert \mu \vert < 4 t)$,
and one can always find zeros of the quasiparticle excitation spectrum
given by $E_{{\bf k}, 2} = 0$. However, the transition from $d$- to
$(se \pm id)$-wave is also very exotic
as it involves a change in the excitation spectrum from
gapless to fully gapped with a corresponding change in topology
of the quasiparticle-quasihole excitation manifold, and
a change in order parameter symmetry accompanied by
the spontaneous breaking of time-reversal.

Since in standard condensed matter physics the interactions
in the same material can not be tunned, one can hope to visit
different phases by changing the temperature or tuning the filling
factor (carrier density)~\cite{goldman-2006}.
If the tuning of carrier density via electrostatic means can be achieved
experimentally for complex oxides, then quantum phase transitions
may be studied as a function of filling factor~\cite{ahn-2006}.
Since electrostatic tuning of carrier density can only be implemented in
thin films or at the surface of bulk materials, additional experiments
to characterize the various phases are difficult. However, in such geometry,
measurements of the penetration depth $\lambda (\nu, T)$ may be possible.
Given that $\lambda (\nu, T) \propto \rho^{-2} (\nu, T)$, where $\rho (\nu, T)$
is the superfluid density, then a distinguished
low-temperature behavior varying from a linear increase with temperature
in the $d$-wave phase to an exponentially activated behavior
in the $se$-wave phase could be revealed as the filling factor $\nu$ is varied.

\begin{figure} [htb]
\centerline{\scalebox{0.50}{\includegraphics{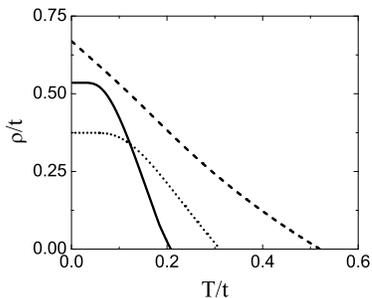} } }
\caption{ \label{fig:superfluid-density} The superfluid density
$\rho/t$ versus temperature $T/t$ at fixed interaction $V/t = 3.0$.
The dotted line corresponds to $se$-wave for  $\nu = 0.24$, the
solid line to $(se \pm id)$-wave for $\nu = 0.38$, and the dashed
line to $d$-wave for $\nu = 0.60$.}
\end{figure}

In order to describe the behavior of $\lambda (\nu, T)$, we
calculate the superfluid
density tensor
\begin{equation}
\label{eqn:superfluid-density} \rho_{ij} (\nu, T) = \frac{1}{L^2}
\sum_{\bf k} \left[ 2 n_{\bf k} \partial_i \partial_j \xi_{\bf k} -
Y_{\bf k} \partial_i \xi_{\bf k} \partial_j \xi_{\bf k} \right],
\end{equation}
at low $T$, where $ n_{\bf k} = (1/2)\left[ 1 - \xi_{\bf k} \tanh (
E_{{\bf k}, 2}/2T )/E_{ {\bf k}, 2} \right] $ is the momentum
distribution, and $Y_{\bf k} = (2T)^{-1} {\rm sech}^2 ( E_{{\bf k},
2}/2T )$ is the Yoshida function. Notice that $\rho_{xx} =
\rho_{yy}$ and $\rho_{xy} = \rho_{yx}$ for all order parameter
symmetries. In Fig.~\ref{fig:superfluid-density}, we show the
temperature dependence of the superfluid density in the $se$-, $(se
\pm id)$-, and $d$-wave phases. At low temperatures, the superfluid
density for the $se$-wave phase exhibits an exponentially activated
behavior $\rho_{se} (0) - \rho_{se} (T)  \sim \exp ( -\vert \tilde
\Delta_{se} \vert / T )$, due to the presence of full gap in the
quasiparticle excitation spectrum $E_{{\bf k},2}$, with similar
behavior for the $(se \pm id)$ phase. However, in the $d$-wave case
the superfluid density decreases linearly with temperature $\rho_d
(0) - \rho_{d} (T)  \sim  T$, as expected from the nodal structure
of $E_{{\bf k}, 2}$.

Before concluding, we would like to make an important remark.
For the model described here there is no
antiferromagnetic phase near half-filling.
This phase emerges by replacing the on-site attractive-$U$ term by
an on-site repulsive (Hubbard-$U$) term in the Hamiltonian
of Eq.~(\ref{eqn:hamiltonian}). The full solution of the
problem including the on-site repulsion $U$, and the nearest neighbor
attraction $(-V)$ is quite complex, however one can make a few qualitative
statements for $U/t \gg V/t$ and fixed $V/t$.
In this case, the system is an antiferromagnetic insulator at and near
half-filling $(\nu = 1)$, however away from it the effects of a locally
repulsive term are dramatically reduced and with decreasing
filling factor superconductivity is achieved first for the $d$-wave phase,
then for the mixed phase $(se \pm id)$ and finally for the $se$-wave phase.
Since the change in symmetry of the order parameter from $d$- to
mixed phase to $se$-wave occurs reasonably far away from half-filling,
such a transition is not dramatically affected by $U$. Thus, even in
more realistic models for complex oxides
such as the cuprates,
the transition proposed here should persist at lower filling
factors.

%
%

We have discussed a simple extended attractive Hubbard model to
describe single-band complex oxide superconductors in
two-dimensions, where the filling factor can be adjusted via
electrostatic doping. Based on symmetry alone, we established that
the possible superconducting states are extended $s$-wave $(se)$,
$d$-wave $(d)$ and mixed phases which break $(se \pm i d)$ and do
not break $(se \pm d)$ time-reversal symmetry. However, we found
that only the $se$-, $d$- and $(se \pm id)$ phases are accessible
within a nearest neighbor attraction model, and that there exists a
tetracritical point where the normal and all superconducting phases
meet at finite temperature. We have shown that quantum phase
transitions between various superconducting phases take place at
filling factors far from half-filling, and analysed the temperature
dependence of the superfluid density (penetration depth) near the
boundaries of such transitions, where the characteristic power law
behavior of the $d$-wave symmetry is replaced by the exponentially
activated behavior of the $se$-wave symmetry.

\acknowledgements{We thank Allen Goldman for discussions
and NSF (DMR-0709584) for support.}

\end{document}